\documentstyle[aps,prd,eqsecnum,epsf]{revtex}  % journal submission

\def\qbar{{\overline{q}}}
\def\half{{\frac{1}{2}}}
\def\pslash{{p \!\!\!/}}
\def\hatZ{{\hat Z}}
\def\lcm{{m}}
\def\lcq{{q}}

\def\gtwid{\raise.3ex\hbox{$>$\kern-.75em\lower1ex\hbox{$\sim$}}}
\def\ltwid{\raise.3ex\hbox{$<$\kern-.75em\lower1ex\hbox{$\sim$}}}

\def\ie{{\it i.e.},\ }
\def\eg{{\it e.g.},\ }
\def\et{{\it et al.}}

\def\cC{{\cal C}}

\def\cJ{{\cal J}}
\def\cK{{\cal K}}

\begin{document}

\draft  
% \draft command makes pacs numbers print

%
% draft title
%
\title{A calculation of the Lepage-Mackenzie scale for the lattice
axial and vector currents}

\author{Claude Bernard and Maarten Golterman}
\address{
Department of Physics, Washington University, 
St.~Louis, Missouri 63130, USA
}
\author{Craig McNeile}
\address{
Department of Physics, University of Utah, 
Salt Lake City, Utah 84112, USA
}
%
%
%%%%%%%%%%%%%%%%%%%%%%%%%%%%%%%%%%%%%%%%%%%%%%%%%%%%%%%%%%%%%%%%%%%%%%

\maketitle

\begin{abstract}

We calculate the perturbative scales ($q^\star$ )  
for the axial and vector currents 
for the Wilson action, with and without tadpole improvement, using
Lepage and Mackenzie's formalism. 
The scale for the pseudoscalar density (times the mass) is
computed as well. Contrary to naive expectation,
tadpole improvement reduces $q^\star$ by only a small amount for 
the operators we consider. We also discuss the use of a
nonperturbative coupling to calculate the perturbative scale.

\end{abstract}


\section{Introduction} 

The calculation of weak matrix elements, such as the 
$f_B$ decay constant, is one of the most important applications of
lattice QCD, because it is this nonperturbative physics that
obscures possible physics beyond the standard model. Although lattice
QCD offers a nonperturbative method of calculating weak matrix
elements from first principles, in practice a perturbative
calculation is also required to extract the continuum number.
Lattice perturbation theory is much harder than continuum perturbation
theory, because there is less symmetry on the lattice that
can be used to simplify expressions. This means that 
lattice perturbative quantities are typically only known to one loop in
perturbation theory. It seems important to extract as much information
from the one-loop calculation as possible.

In the past, one-loop
lattice perturbation theory was considered unreliable,
because for many quantities the one-loop corrections are large
if one uses the bare lattice coupling constant,
and the agreement with results that could be calculated 
numerically was the very poor. Lepage and Mackenzie~\cite{Lepage93} 
then showed that the use of a more ``physical" coupling, calculated at
a judiciously chosen scale (generically referred to as $q^\star$) 
and the resummation of gluon-tadpole contributions to the 
perturbative expansion improve the situation for many quantities.

In this paper we calculate the scale $q^\star$ using the Lepage and
Mackenzie prescription for the local
axial and vector currents (as well as the mass times the
pseudoscalar density), with and without tadpole improvement.
Some of our results were also obtained
in Ref.~\cite{Crisafulli:1997ic};
however that paper did not study tadpole-improved perturbation
theory.  In addition, the current work shows how to deal with
certain ambiguous (though numerically small) contributions to the
calculation in Ref.~\cite{Crisafulli:1997ic}.

There are a number  of nonperturbative renormalization
techniques that have been
used in numerical 
computations~\cite{Martinelli94a,Vladikas:1995qt,Luscher:1997jn}. These
techniques are very interesting, but they have not as yet been widely used in
large scale simulations.  The calculation of the nonperturbative
renormalization constants required in the calculation of $f_B$ in the
simulations by the MILC collaboration~\cite{BernardfB98}, for
example, would require, at the least, a substantial investment in computer 
time.
There are also conceptual problems in generalizing existing 
nonperturbative methods to
the case of large quark mass.
Furthermore, it is an important consistency
check on the numerical calculation of renormalization factors,
that they should agree with the weak-coupling results at small enough
coupling, because there are potential systematic errors, such as
lattice-spacing artifacts, in the 
numerical results.
This motivates trying to extract as much information as possible from
lattice perturbation theory calculations.

The outline of this paper is as follows: in section 2 we calculate the
renormalization constants from Ward identities, while in section 3, they
are calculated by matching on-shell matrix elements to the continuum.
In section 4, these renormalization constants are tadpole improved, and
in section 5, we calculate the scales $q^\star$ for each of them with and
without tadpole improvement.  In section 6, we discuss the stability of
our results under changes in the scale-setting prescription.  We end
by providing some insight into the fact why tadpole improvement does
not lower the scale $q^\star$ for these quantities as much as one might
naively expect.

\section{\bf $Z_A$, $Z_{\lcm P}$ and $Z_V$ from Ward-identities}
\label{se:WARD-ID}

First, we will derive the renormalization factors from the
Ward identities~\cite{COLLINS,Trueman:1979en}.
We consider the following Ward identity (WI) for the nonsinglet flavor
axial current $A_\mu=A^a_\mu T^a$ ($T^a$ are the hermitian, traceless
flavor generators; we work in euclidean space):
\begin{eqnarray}
\langle\sum_\mu\partial_\mu A^a_\mu(x)q(z)\qbar(y)\rangle&=&
2m\langle P^a(x)q(z)\qbar(y)\rangle \label{wi} \\
&&-\langle q(z)\qbar(y)
\rangle T^a\gamma_5\delta(x-y)-T^a\gamma_5\langle q(z)\qbar(y)
\rangle\delta(x-z)\;, \nonumber
\end{eqnarray}
with
\begin{eqnarray}
A^a_\mu(x)&=&\qbar(x)\gamma_\mu\gamma_5 T^a q(x)\;, \label{ac} \\
P(x)&=&\qbar(x)\gamma_5 T^a q(x)\;. \label{ps}
\end{eqnarray}
This WI holds for any regulator which respects the (nonsinglet) axial
flavor symmetry, such as naive dimensional regularization (NDR), in
which $\gamma_5$ anticommutes with $\gamma_\mu$ for all $\mu=1,\dots,d$.
Since $A^a_\mu$ is the only dimension-three axial vector operator that
can be built out of quark and gluon fields, it does not mix with other
operators.  Taking $m=0$ in eq.~(\ref{wi}), 
we see that $A^a_\mu$ does not require renormalization
(the RHS of eq.~(\ref{wi}) is finite
after renormalizing the quark fields $q(z)$ and $\qbar(y)$). 
Thus, if $Z_A$ is defined to be the renormalization constant for $A^a_\mu$
which preserves the WI~(\ref{wi}), we have $Z_A=1$ in any scheme, such
as NDR, which preserves the axial flavor symmetry.
The operators $mP^a$ also do not
mix with any other operators, and therefore $Z_{mP}$ (defined to be the
renormalization constant of the operator $mP^a$) may be determined by
considering the case $m\ne 0$ after $Z_A$ has been defined.  In NDR, we
have $Z_{mP}=1$. 

Wilson fermions break the axial flavor symmetry, and no conserved axial
currents exist, so that the WI introduced above does not hold for any
axial current on the lattice.  Here, we are interested in the local 
(\ie\ single-site)
axial current and pseudo-scalar density as defined in 
eqs.~(\ref{ac},\ref{ps}), 
and we would like to find the multiplicative renormalization 
constants $Z_A$ and $Z_{mP}$ such that eq.~(\ref{wi}) holds in the scaling 
region:
\begin{eqnarray}
Z_A\langle\sum_\mu\partial_\mu A^a_\mu(x)q(z)\qbar(y)\rangle&=&
2Z_{mP}(M-M_c)\langle P^a(x)q(z)\qbar(y)\rangle \label{eq:wicl} \\
&&-\langle q(z)\qbar(y)
\rangle T^a\gamma_5\delta(x-y)-T^a\gamma_5\langle q(z)\qbar(y)
\rangle\delta(x-z)\;. \nonumber
\end{eqnarray}
Here $M=m+4r$ ($r$ is the Wilson parameter and $m$ is the bare mass) is
the coefficient of the single-site term in the Wilson--Dirac lagrangian, 
and $M_c$ is its critical value.
On the lattice, we may write
\begin{eqnarray}
\langle\sum_\mu\Delta_\mu A^a_\mu(x)q(z)\qbar(y)\rangle&=&
2(M-M_c)\langle P^a(x)q(z)\qbar(y)\rangle \label{wib} \\
&&-\langle q(z)\qbar(y)
\rangle T^a\gamma_5\delta(x-y)-T^a\gamma_5\langle q(z)\qbar(y)
\rangle\delta(x-z) \nonumber \\
&&+\langle E^a(x)q(z)\qbar(y)\rangle\;, \nonumber 
\end{eqnarray}
where $\Delta_\mu$ is the backward difference operator 
$\Delta_\mu f(x)=f(x)-f(x-\mu)$, and $E^a$ is the ``evanescent" operator  
\begin{eqnarray}
E^a(x)&=&\sum_\mu\Delta_\mu A^a_\mu(x)-2(M-M_c)P^a(x) \label{eo} \\
&&+\qbar(x)\gamma_5 T^a(D+M)q(x)+\qbar(x)({\buildrel\leftarrow\over
D}+M)\gamma_5 T^a q(x)\;, \nonumber
\end{eqnarray}
in which
\begin{equation}
Dq(x)=\half\sum_\mu\left[(\gamma_\mu-r)U_\mu(x)q(x+\mu)
-(\gamma_\mu+r)U^\dagger_\mu(x-\mu)q(x-\mu)\right] \label{Ddef} 
\end{equation}
($D+M$ is the Wilson--Dirac operator).

Going to momentum space by multiplying eq.~(\ref{wib}) by
$\sum_{xyz}{\rm exp}(-ikx-ipz+ip'y)$, and amputating the external 
quark lines, we find, at one-loop order in $g$,
\begin{equation}
\langle E^a(k)q(p)\qbar(p')\rangle_{\rm amp}^{\rm one-loop}
=C_F g^2\delta(p+k-p')T^a I(p,p')\;,
\label{E1pi} 
\end{equation}
with $C_F=\frac{N^2-1}{2N}$ the quadratic Casimir for $SU(N)$, and
\begin{eqnarray}
I(p,p')&=&-\sum_\nu(1-e^{i(p-p')_\nu})\sum_\mu\int_\ell D(\ell)\times
\label{I} \\
&&\phantom{-\sum_mu\int_\ell D(\ell)\Biggl(}
\Biggl(
\Gamma^-_\mu S(p+\ell)\gamma_\nu\gamma_5 S(p'+\ell)\Gamma^-_\mu\;e^{i(p+p'+\ell)_\mu}
\nonumber \\
&&\phantom{-\sum_mu\int_\ell D(\ell)\Biggl(}
+\Gamma^-_\mu S(p+\ell)\gamma_\nu\gamma_5 S(p'+\ell)\Gamma^+_\mu\;e^{i(p-p')_\mu}
\nonumber \\
&&\phantom{-\sum_mu\int_\ell D(\ell)\Biggl(}
+\Gamma^+_\mu S(p+\ell)\gamma_\nu\gamma_5 S(p'+\ell)\Gamma^-_\mu\;e^{-i(p-p')_\mu}
\nonumber \\
&&\phantom{-\sum_mu\int_\ell D(\ell)\Biggl(}
+\Gamma^+_\mu S(p+\ell)\gamma_\nu\gamma_5 S(p'+\ell)\Gamma^+_\mu\;e^{-i(p+p'+\ell)_\mu}
\Biggr)
\nonumber \\
&&+2(M-M_c)\sum_\mu\int_\ell D(\ell)\Biggl(
\Gamma^-_\mu S(p+\ell)\gamma_5 S(p'+\ell)\Gamma^-_\mu\;e^{i(p+p'+\ell)_\mu}
\nonumber \\
&&\phantom{-\sum_mu\int_\ell D(\ell)\Biggl(}
+\Gamma^-_\mu S(p+\ell)\gamma_5 S(p'+\ell)\Gamma^+_\mu\;e^{i(p-p')_\mu}
\nonumber \\
&&\phantom{-\sum_mu\int_\ell D(\ell)\Biggl(}
+\Gamma^+_\mu S(p+\ell)\gamma_5 S(p'+\ell)\Gamma^-_\mu\;e^{-i(p-p')_\mu}
\nonumber \\
&&\phantom{-\sum_mu\int_\ell D(\ell)\Biggl(}
+\Gamma^+_\mu S(p+\ell)\gamma_5 S(p'+\ell)\Gamma^+_\mu\;e^{-i(p+p'+\ell)_\mu}
\Biggr)
\nonumber \\
&&-\Sigma(p,m)\gamma_5-\gamma_5\Sigma(p',m)\;, \nonumber \\ 
\Sigma(p,m)&=&
\sum_\mu\int_\ell D(\ell)\Biggl(
-\Gamma^-_\mu S(p+\ell)\Gamma^-_\mu\;e^{i(2p+\ell)_\mu}
-\Gamma^+_\mu S(p+\ell)\Gamma^+_\mu\;e^{-i(2p+\ell)_\mu}
\nonumber \\
&&\phantom{+\sum_\mu\int_\ell D(\ell)\Biggl(}
-\Gamma^-_\mu S(p+\ell)\Gamma^+_\mu
-\Gamma^+_\mu S(p+\ell)\Gamma^-_\mu 
\label{se} \\
&&\phantom{+\sum_\mu\int_\ell D(\ell)\Biggl(}
+\frac{i}{2}\sin{p_\mu}\gamma_\mu-\frac{r}{2}\cos{p_\mu}\Biggr)\;.
\nonumber 
\end{eqnarray}  
Here $\Sigma(p,m)$ is the one-loop fermion self-energy 
(defined to be equal to the sum of diagrams, with no overall minus sign),
and $S(p)=
\left[i\sum_\mu\gamma_\mu\sin{p_\mu}+m+r\sum_\mu(1-\cos{p_\mu})\right]^{-1}$
is the tree-level fermion propagator.  The gluon propagator in the Feynman
gauge is $\delta_{\mu\nu}D(p)$ with
$D^{-1}(p)=4\sum_\mu\sin^2{(p_\mu/2)}$.  The matrices $\Gamma^\pm_\mu$ are
defined by
\begin{equation}
\Gamma^\pm_\mu=\half(\gamma_\mu\pm r)\;. \label{Gammas}
\end{equation}
Furthermore, we abbreviated $\int_\ell\equiv\frac{1}{(2\pi)^4}
\int_{|\ell_\mu|\le\pi}d^4 \ell$.

We now observe that $I(0,0)=-2\gamma_5\Sigma(0,0)$ provides the counterterm
to remove the $1/a$ divergence in the contact terms in eq.~(\ref{wib}) (after
amputation of external fermion lines).
A straightforward calculation then shows that the remainder,
$I(p,p')-I(0,0)$, is finite, and we can take the continuum limit.
We expand to linear order in $p$, $p'$ and
$m=M-M_c+O(g^2)$: 
\begin{equation}
C_F \left(I(p,p')-I(0,0)\right)
=c_A\;i(\pslash'-\pslash)\gamma_5-2c_{mP}\;m\gamma_5\;, \label{Icl} 
\end{equation}
from which we then obtain the renormalization constants
\begin{equation}
Z_A=1-g^2c_A\;,\ \ \ \ \ Z_{mP}=1-g^2c_{mP}\;,
\label{Zresult}
\end{equation}
with, for $N=3$ and $r=1$,
\begin{equation}
c_A=0.133373(2)\;,\ \ \ \ \ c_{mP}=0.081419(2)\;. \label{zazp}
\end{equation}

At this point, we pause to comment on this derivation.  Generally, in
calculating the continuum limit of lattice integrals such as $I(p,p')$,
special care is required near the origin in momentum space.  Near $\ell=0$
in eq.~(\ref{I}), one may approximate the integrand by its covariant
form.  In this case, with the routing of the external momenta through
the fermion line, it turns out that the covariant form of the integrand
vanishes identically (as explained below), and 
therefore eq.~(\ref{Icl}) is obtained by a
straightforward expansion in $p$, $p'$ and $m$.  This would not be true
had we chosen the routing of the external momenta through the gluon
line.  This will turn out to be relevant for the definition of
$q^\star$, and will be discussed further below. 

A completely analogous analysis may be performed in order to find the
renormalization constant $Z_V$ of the local vector current
\begin{equation}
V^a_\mu(x)=\qbar(x)\gamma_\mu T^a q(x)\;. \label{V}
\end{equation}
For Wilson fermions, this current is not conserved, and we find
\begin{equation}
Z_V=1-g^2 c_V\;,\ \ \ \ \ c_V=0.174083(3)\;. \label{ZV}
\end{equation} 
Equations~(\ref{zazp},\ref{ZV}) are in agreement with previous computations
\cite{Martinelli:1983mw,Meyer:1983ds,Groot:1984ng}.

\section{\bf $Z_A$, $Z_{\lcm P}$ and $Z_V$ by on-shell matching to the continuum}
\label{se:MATCHING}

A common approach in the literature is to compute the
renormalization constants by matching an on-shell lattice perturbative computation
to the continuum, where the renormalization constants are known.  
We present this approach here because the value of $q^\star$ 
depends on the specific form of the
integrands. It is therefore
important to check that a separate derivation of the integrals
will produce a consistent result for $q^\star$.  

For definiteness,
we focus first on the nonsinglet axial current, $A_\mu$.  The 
derivations for the vector
($V_\mu$) and pseudoscalar ($mP$) cases are similar, but there
are a few important differences, which we note explicitly.
Because $A_\mu$ has no anomalous dimension or mixings, we may
write its one-loop matrix element  between on-shell quark states
$i$ and $f$ in
regularization scheme $S$ as
\begin{equation}
\langle f \mid (A^a_\mu)^S \mid i \rangle  = 
\gamma_\mu \gamma_5 T^a
(1 + g^2 \cC_A^S (m,\lambda)) \ ,
\label{eq:MATRIXELEMENT}
\end{equation}
where $m$ is the quark mass,
$\lambda$ is a gluon mass that has been inserted to regularize
the infrared divergences, and spinors on the external lines are implicit.

We then define a on-shell renormalized axial current 
$(A^a_\mu)^{\rm shell}$
\begin{eqnarray}
(A^a_\mu)^{\rm shell} & \equiv & \tilde Z^S_A(m,\lambda) (A^a_\mu)^S \\
\tilde Z_A^S(m,\lambda) & = &
 1 - g^2 \cC^S_A (m,\lambda) +\dots\ .
\label{eq:SHELL}
\end{eqnarray}
By definition, the on-shell quark matrix element of $(A^a_\mu)^{\rm shell}$ 
is the same in any scheme and equal to $\gamma_\mu \gamma_5 T^a$, as
 long as $m$ and $\lambda$ 
are the same as in eq.~(\ref{eq:SHELL}).
 Because there is no operator mixing, this then implies the 
scheme independence of
all renormalized Green's functions with a $(A^a_\mu)^{\rm shell}$ insertion.
However, we may also define a renormalized current 
$(A^a_\mu)^{\rm WI}$ using the
Ward Identity, 
eq.~(\ref{eq:wicl}):
\begin{equation}
(A^a_\mu)^{\rm WI}  \equiv  Z^S_A\;(A^a_\mu)^S \ ,
\label{eq:WI}
\end{equation}
where the superscript $S$ specifies the scheme to which eq.~(\ref{eq:wicl})
is applied.  
$Z^S_A$ is UV finite; it is free of infrared divergences since it is 
defined off-shell. By dimensional analysis,
$Z^S_A$ is therefore also independent of the quark mass $m$.
Since renormalized Green's functions 
with a $(A^a_\mu)^{\rm WI}$ insertion are also scheme independent,
we have, for any two schemes $S$ and $S'$,
\begin{equation}
{Z_A^S\over 
\tilde Z_A^S(m,\lambda)} =
{ Z_A^{S'} \over
\tilde Z_A^{S'}(m,\lambda)}
\label{eq:ZRATIOS}
\end{equation}

We now specialize to the case $S={\rm lattice}$ and $S' = {\rm NDR}$. 
As long as the regulator does not violate the symmetry, 
the Lie algebra of conserved charges  fixes the on-shell matrix
elements of the corresponding currents, guaranteeing that a conserved
nonabelian flavor current  is not renormalized.  Therefore
\begin{equation}
\tilde Z_A^{\rm NDR}(m=0,\lambda) = 1
\label{eq:ZANDR}
\end{equation}
since the nonsinglet axial current is conserved when $m=0$.
Similarly,
\begin{equation}
\tilde Z_V^{\rm NDR}(m,\lambda) = 1
\label{eq:ZVNDR}
\end{equation}
for {\em all}\/ $m$, $\lambda$.  
A simple computation shows, however, that
$\tilde Z_A^{\rm NDR}(m,\lambda) \not= 1$ 
for general $m, \lambda$,  
(including the limit $\lambda\to0$, $m$ fixed).
Note that the presence of the gluon mass
in eqs.~(\ref{eq:ZANDR}, \ref{eq:ZVNDR}) does not violate the flavor
symmetries, although  it
would violate gauged symmetries. 

It is useful to see explicitly how eqs.~(\ref{eq:ZVNDR}) 
and (\ref{eq:ZANDR}) arise at one-loop. At this order, the computations
are identical to those in QED, and amount the statement
that $Z_1 = Z_2$ \cite{BJandD}.  If one 
routes the external momentum  $p$ through the fermion line, 
then the cancellation of the wave-function and vertex renormalizations
which results in eq.~(\ref{eq:ZVNDR}) can easily be shown from the
(Euclidean) identity
\begin{equation}
{\partial\over \partial p^\mu} {1\over (i \not\! p + m) } = 
{1\over (i \not\! p + m) } (-i \gamma_\mu) {1\over (i \not\! p + m) }
\label{eq:PROP-DERIV}
\end{equation}
Equation~(\ref{eq:PROP-DERIV}), coupled with the on-shell definition of the
wave-function renormalization,
\begin{equation}
Z_2 =  1 + {p^\mu\over m} 
{\partial \Sigma \over \partial p^\mu}\biggr\vert_{ip\!\!\!/= -m}\ ,
\label{eq:Z2}
\end{equation}
immediately gives $Z_1 = Z_2$.  With the chosen momentum routing,
the equality holds at the level of the Feynman integrands. 
On the other
hand, if the external momentum is routed through the gluon line,
there is no obvious relation between the integrands for $Z_1$ and $Z_2$.
The equality then appears only after the momentum and Feynman parameter
integrals are performed and requires a regulator which permits 
shifts of the loop momentum without the generation of boundary terms.
This is the reason that the ``covariant part'' of the lattice integral
in Section~\ref{se:WARD-ID} vanishes only when the external
momentum is routed through the fermion line.  Although complete
lattice integrals are invariant under shifts, the ``covariant part'' is
not because it is the integral of a continuum-like integrand over
a sphere of radius $\delta$ (the ``inner region'' of 
Ref.~\cite{KARSTENandSMIT}) or of radius $\pi$ 
(the integration region chosen for the
continuum-like integrals added and subtracted in Ref.~\cite{Bernard87a}).
The boundary term generated by a shift in the loop momentum 
thus does not vanish in the continuum
limit for a linearly divergent integral such as the self-energy.

For $mP$, one can relate the mass derivative of the self-energy
to the vertex function (at least at one loop), 
and again show at the level of the
Feynman integrands (with the proper momentum routing) that
\begin{equation}
\tilde Z_{mP}^{\rm NDR}(m=0,\lambda) = 1\ .
\label{eq:ZPNDR}
\end{equation}

Returning to the discussion of $Z_A^{\rm lattice}$,
we combine eqs.~(\ref{eq:ZRATIOS}) and (\ref{eq:ZANDR}), and use
$Z^{\rm NDR}_A=1$ (since NDR preserves the chiral symmetry), giving
\begin{equation}
Z_A^{\rm lattice} = \tilde Z_A^{\rm lattice}(0,\lambda) \ .
\label{eq:ZALIMIT}
\end{equation}
The corresponding relation holds for $Z_V^{\rm lattice}$ 
and $Z_{mP}^{\rm lattice}$.
Note that all direct reference to the continuum has been eliminated.
Since from now on we talk only about lattice quantities we can drop
the qualifier ``lattice'' from eq.~(\ref{eq:ZALIMIT}) and return
to the notation of Section \ref{se:WARD-ID}.
We remark that the renormalization factors were computed 
in Ref.~\cite{Groot:1984ng} using the equivalent of
eq.~(\ref{eq:ZALIMIT}). 

The computation of the right hand side of eq.~(\ref{eq:ZALIMIT})
is fairly easy, since it can be done in the massless limit.
We need to compute the self-energy graphs (the continuum-like
graph and the lattice tadpole)
and the vertex correction graph (for each current).
To avoid computing any covariant
parts, we route the external momentum  through the fermion line.
This routing will also be very convenient in the computation of $q^\star$.

We write the lattice self-energy as
\begin{equation}
\Sigma(p) = {1\over a}\Sigma_0(p^2,m^2,\lambda^2) + i\!\not\! p\;\Sigma_1(p^2,m^2,\lambda^2) + m\Sigma_2(p^2,m^2,\lambda^2) \ .
\label{eq:SELFENERGYDEF}
\end{equation}
Only $\Sigma_1$ is needed for wave function renormalization at $m=0$;
$\Sigma_2$ is needed for $Z_{mP}$.
Computing the self-energy graphs with the methods of Ref.~\cite{Bernard87a},
we have
\begin{equation}
\Sigma_{1,2}(0,0,\lambda^2) = g^2 C_F \int_\ell
I_{1,2}\ .
\label{eq:SIGMAINTEGRAL}
\end{equation}
Taking Wilson $r=1$, the integrands $I_{1,2}$ are:
\begin{eqnarray}
I_1  & = & {1\over 8\Delta_1} + {1\over 4\Delta_1\Delta_2}\left(-2 + {11\Delta_1
\over 2} - 2\Delta_1^2 - {\Delta_4\over 2} 
+ {1\over \Delta_2}\left(\Delta_4 -\Delta_5 + \Delta_1\Delta_4 +
4\Delta_1\Delta_5 - 2\Delta_1^2 \Delta_4 - 2\Delta_{13}\right)\right) + \cr
 & & \quad + {\theta(\pi^2 - \ell^2)\over \ell^4} \  + covariant\ part,\cr
I_2 & = & - {4\Delta_1(\Delta_1-2) + \Delta_4 \over \Delta_2^2}  + {\Delta_1- 2
\over 2\Delta_1\Delta_2} 
 + {4\theta(\pi^2 - \ell^2)\over \ell^4} \  + covariant\  part,
\label{eq:I12}
\end{eqnarray}
where $\Delta_1 \dots \Delta_5$ are defined in 
Refs.~\cite{Bernard87a,Martinelli84a} (for $r=1$):
\begin{eqnarray}
\Delta_{1}  &=&  \sum_\alpha \sin^2({\ell_\alpha\over 2})\;,\cr
\Delta_{4}  &=&  \sum_\alpha \sin^2(\ell_\alpha)\;,\cr
\Delta_{5}  &=&  \sum_\alpha \sin^2({\ell_\alpha\over 2})
\sin^2(\ell_\alpha)\;,\cr
\Delta_2  &=& \Delta_4 + 4 \Delta_1^2\;,
\label{DELTAS}
\end{eqnarray}
and we introduce
\begin{equation}
\Delta_{13}  \equiv  \sum_\alpha \sin^4({\ell_\alpha\over 2})
\sin^2(\ell_\alpha)\;.
\label{eq:DEL13}
\end{equation}
The ``covariant parts''  in eq.~(\ref{eq:I12})
do not 
need to be specified further, since they will cancel in the final
answers, as explained above.  The terms proportional  to $1/\ell^4$,
which come from the expansion in powers of $a$
of added and subtracted covariant integrals,
are written explicitly here, since they are needed
to cancel the logarithmic divergences in the
rest of the integrals.  However, these terms will also cancel in the 
final results.

In Ref.~\cite{Bernard87a}, the external momentum in the self-energy computation
was chosen to flow through the gluon line.
(This saved some effort in the expansion in powers of $a$, since the
fermion propagator is more complicated than the gluon one.)  
For our current purposes,
we therefore have performed the calculation of $I_1$ from scratch.
In the case of $I_2$, however,  
one may set $p=0$ from the beginning, so that the
routing of the external momentum is irrelevant.  Then $I_2$ may be
taken immediately from Ref.~\cite{Bernard87a}.

For the vertex diagrams we may also set $p=0$ {\it ab initio} and
take the results from Ref.~\cite{Bernard87a}.
We have 
\begin{equation}
\Lambda_{\gamma_\mu\gamma_5,\gamma_\mu,\gamma_5}  
= (\gamma_\mu\gamma_5,\gamma_\mu,\gamma_5)g^2 C_F 
\int_\ell 
\cJ_{\gamma_\mu\gamma_5,\gamma_\mu,\gamma_5}\;,
\label{VERTEXDEF}
\end{equation}
where
\begin{eqnarray}
\cJ_{\gamma_\mu\gamma_5}  & = &  4\cK'_2 + 2\cK_1 + \cK_0\;,\cr
\cJ_{\gamma_\mu}  & = &  4\cK'_2 - 2\cK_1 + \cK_0\;,\cr
\cJ_{\gamma_5}  & = &  16\cK'_2 -4\cK_1 + \cK_0\;, 
\label{eq:JDEF}
\end{eqnarray}
with
\begin{eqnarray}
\cK'_2  & = & {1\over 48\Delta_2^2}\left({\Delta_5
\over \Delta_1} - \Delta_4 \right)
+  {\Delta_4 \over 16 \Delta_1 \Delta_2^2} 
 -{\theta(\pi^2 - \ell^2)\over 4\ell^4}  + covariant\ part,\cr
\cK_1  & = & {1\over 16\Delta_2^2}\bigl(\Delta_4 + 4
\Delta_1(\Delta_1 -4)\bigr)\;,\cr
\cK_0  & = & {1\over 12\Delta_2^2}\left(\Delta_4 - 4{\Delta_5\over\Delta_1}
+6\Delta_4 + 12\Delta_1^2\right) \ .
\label{eq:KAYS}
\end{eqnarray}

If we now write the $Z$ factors in terms of the constants
$c_A$, $c_V$ and $c_{mP}$ as in eqs.~(\ref{Zresult},\ref{ZV}),
we have
\begin{eqnarray}
c_A & = & C_F \int_\ell
(I_1 + \cJ_{\gamma_\mu\gamma_5})\;, \cr
c_V & = & C_F \int_\ell
(I_1 + \cJ_{\gamma_\mu})\;, \cr
c_{mP} & = & C_F \int_\ell 
(I_2 + \cJ_{\gamma_5}) \;.
\label{eq:CINTEGRALS}
\end{eqnarray}
Explicit calculation shows that the  covariant parts of 
$I_{1,2}$ and $\cJ_{\gamma_\mu\gamma_5,\gamma_\mu,
\gamma_5}$ cancel in  eq.~(\ref{eq:CINTEGRALS}), as expected from
the general arguments above.
Numerical integration of eq.~(\ref{eq:CINTEGRALS}) then reproduces
eqs.~(\ref{zazp},\ref{ZV}).

\section{Tadpole improvement } 
\label{se:tadpole}

The idea behind tadpole improvement~\cite{Lepage93}
is based on the observation that,
for the relatively large values of the lattice coupling constant $g$
used in current lattice QCD computations, the expectation value of the
link variable, $u_0\equiv\langle U_{\mu x}\rangle$ (defined, \eg\ in
Landau gauge), is not very close to one.  If one would calculate, 
for instance, the quark propagator in a mean-field approximation by
replacing $U_{\mu x}\to u_0$ in the quark action, one would find 
\begin{equation}
\langle q(x)\qbar(y)\rangle_{\rm MF}=u_0^{-1}
\langle q(x)\qbar(y)\rangle_{\rm free}(M\to u_0^{-1}M)\;,\label{mfprop}
\end{equation}
where the notation $M\to u_0^{-1}M$ indicates that we replace $M$ by
$u_0^{-1}M$ in the expression for the free Wilson propagator.

This implies that, in order to make contact with the continuum, the
quark field has to be renormalized by a factor $\sqrt{u_0}$.  For local
bilinears like $A_\mu^a(x)$, $P^a(x)$ and $V_\mu^a(x)$, this results in
a factor $u_0$:
\begin{equation}
A_\mu^a(x)_{\rm cont}=u_0 A_\mu^a(x)_{\rm lattice}\;,\label{cltree}
\end{equation}
and similarly for the other operators.
From the mean-field result eq.~(\ref{mfprop}) for the quark propagator,
we find that $M_c=4ru_0$ and $m_{\rm MF}=u_0^{-1}(M-M_c)$. 
(In terms of the hopping parameter $\kappa=1/(2M)$, we get
$\kappa_c=1/(8ru_0)$.) Therefore,
the operator $(M-M_c)P^a(x)$ does not renormalize at all.

The factor $u_0$ takes into account the gauge-field tadpole diagrams
contributing to the renormalization constants $Z_A$ and $Z_V$. 
Of course, these are not the only contributions.  We may now use 
standard perturbation theory to improve on our mean-field estimates
for the renormalization constants, but we should be careful to take out those
contributions that have already been absorbed into $u_0$~\cite{Bernard94a}.  
It follows that 
\begin{equation}
A_\mu^a(x)_{\rm cont}={{u_0}\over{u_0^{\rm PT}}}Z_A 
A_\mu^a(x)_{\rm lattice}\;,\label{clpt}
\end{equation}
%%%
where $u_0^{\rm PT}$ is the mean link $\langle U_{\mu x}\rangle$
calculated in perturbation theory, $u_0$ is the nonperturbatively
(\ie\ numerically) computed value (using the same definition!), and
$Z_A$ is the usual renormalization factor, calculated to one loop
in eq.~(\ref{Zresult}).  This leads to the definition of the tadpole-improved
perturbative renormalization factor
%%%%
\begin{equation}
\hatZ_A=Z_A/u_0^{\rm PT}\;,\label{Zhat}
\end{equation}
%%%
with a similar definition of $\hatZ_V$.

The mean-field link, $u_0$, can be defined in different ways, and the
idea of tadpole improvement is useful to the extent that the various
definitions agree.  
Recently, Lepage~\cite{Lepage:1997id} and Trottier~\cite{Trottier:1997ce}
have 
advocated the use of the 
Landau mean link as the nonperturbative $u_0$ factor.
In Landau gauge, to one loop, we have (for $N=3$)
%%%%
\begin{equation}
u_0^{\rm PT}=1-0.077466(1)g^2 \ \ \ \ \ ({\rm Landau\ gauge}).\label{ulg}
\end{equation} 
%%%%
We may also calculate the mean link from $M_c$:
\begin{equation}
u_0^{\rm PT}=1-0.108570(2)g^2 \ \ \ \ \ ({\rm from\ }M_c).\label{uMc}
\end{equation}
We may now write
\begin{equation}
\hatZ_i=1-c_i g^2\;,\ \ \ \ \ i=A, V\;,\label{cs}
\end{equation}
and tabulate the numerical values for the constants $c_i$ obtained from
eqs.~(\ref{Zhat}--\ref{cs}).
For $u_0^{\rm PT}$ calculated in Landau gauge or
from $M_c$, we report the values in table~\ref{tb:perturbcontants}.

\section{Results for the Lepage-Mackenzie scale $\lcq^\star$}
\label{se:qstar}

The prescription for obtaining the Lepage-Mackenzie scale is 
to calculate $q^{\star}$ from
\begin{equation}
\log ( (q^\star)^2 ) \equiv 
   \frac{ \int d^4 q f(q) \log(q^2) }
    { \int d^4 q f(q)) }\;,
\label{eq:prescription}
\end{equation}
if the one-loop integral for a particular renormalization constant has
the form $\int d^4 q f(q)$, and
where $q$ is the momentum flowing through the gluon line (see below).
The condition that the momentum flow through the 
gluon line is critical both for the understanding
and the correct application of the Lepage and Mackenzie 
scale-setting procedure.

The Lepage and Mackenzie procedure is based on an older
technique for setting the scale in continuum perturbative expressions
due to Brodsky, Lepage, and Mackenzie (BLM)~\cite{Brodsky83}. 
In the BLM 
procedure, the scale is chosen to remove the leading-order coefficient
of the number of fermion flavors from the perturbative expansion.
Physically this absorbs a class of graphs from the 
fermion contribution to the vacuum polarization into the
running coupling. This boils down to the requirement 
that $q$ in eq.~(\ref{eq:prescription}) is the momentum 
flowing through the gluon propagator for self-energy and
vertex-correction diagrams.

The original BLM scale 
setting procedure is more general than the 
Lepage and Mackenzie prescription (eq.~(\ref{eq:prescription})),
because it can cope with perturbative expressions of operators
with nonzero anomalous dimensions. The connection between the 
Lepage and Mackenzie scale setting procedure and the BLM 
procedure has been 
discussed by Kronfeld~\cite{Kronfeld:1997zc}.

We will now calculate $q^\star$, both with and
without tadpole improvement,
using eq.~(\ref{eq:prescription}).
Let us, for definiteness, consider the case of $Z_A$.  The relevant
integral is given in eq.~(\ref{eq:CINTEGRALS}), 
with $I_1$ and $\cJ_{\gamma_\mu
\gamma_5}$ given by eqs.~(\ref{eq:I12},\ref{eq:JDEF},\ref{eq:KAYS}).
Since the loop momentum $q$ has been chosen from the beginning to be
the gluon momentum (\ie the external momentum is routed through
the fermion line), we just need to 
recompute eq.~(\ref{eq:CINTEGRALS}) with $\log(q^2)$ inserted.

In table~\ref{tb:qstarRES}, we give the values of $q^\star$ for the three 
$Z_i$, and
for the three cases of no improvement, Landau-gauge tadpole
improvement, and $M_c$ tadpole improvement.  
For the case of $Z_{mP}$, we did not include $q^\star$ values from 
tadpole improvement, since the operator $mP^A(x)$ is not affected by it,
as explained in the previous section.  

Note that if we had chosen the loop momentum to be the fermion momentum,
the insertion of $\log(q^2)$ into the integrands would
in general give a different result.  While 
we may always shift the loop momentum 
in the original expressions for the diagrams,
this is not true after we have inserted 
$\log{(q^2)}$.  Indeed,  in the case where $q$ is the fermion momentum,
the integral with $\log{(q^2)}$ inserted is not even convergent. The
covariant parts in that case generate integrals roughly of the type
\begin{equation}
\int d^4 q {a^2p^2\over (q^2 + a^2p^2)^3} \ ,
\label{eq:BADINTEGRAL}
\end{equation}
which approaches a constant as $a\to0$ but diverges logarithmically if
$\log{(q^2)}$ is inserted into the integrand before $a\to0$ is taken.
This tells us that this procedure of determining the 
typical scale $q^\star$ is consistent at one loop only when this scale is the
typical momentum of the gluon propagator.  This is in accordance with the
intuitive arguments underlying the BLM scheme. 

In previous computations of $q^\star$ in the 
static-light case~\cite{Hernandez94a} and in the
Wilson or clover case~\cite{Crisafulli:1997ic}, the loop momentum was
not chosen to be the gluon momentum. The straightforward insertion
of $\log(q^2)$ into the integrals was therefore not possible, because
the ``constant terms'' (the $a\to0$ limit of 
integrals like eq.~(\ref{eq:BADINTEGRAL})) would give divergent results.
These terms then had to be treated in an ad-hoc manner:
Hern\'andez and
Hill~\cite{Hernandez94a} replaced the integrals with constants over
the Brillouin zone; Crisafulli \et\ shifted and  redefined the
integration variables until an integral was found which remained
convergent after insertion of $\log(q^2)$.
Luckily, however, these terms contribute only a small amount
to the final results for $q^\star$.  Therefore the ambiguities in
previous computations make only a small numerical difference.
The results in Ref.~\cite{Crisafulli:1997ic}
for the $q^\star$ of $Z_A$ or $Z_V$ 
differ by about 1\% from the current, unambiguous answers.
In making this comparison, we computed the integrals in 
Ref.~\cite{Crisafulli:1997ic} for ourselves (after correcting
two typographical errors in $I_\Sigma$ in eq.~(139)),
since the quoted
answers ($q^\star = 2.4$ for $Z_V$ and $q^\star = 2.6$ for $Z_A$ without
tadpole improvement)
had too few significant figures to see the difference clearly. 

Since the integrals
which define the $c_i$ (eqs.~(\ref{Icl},\ref{ZV}) or (\ref{eq:CINTEGRALS})) 
are finite
and represent cutoff effects, one would anticipate $q^\star$ for these
quantities to be of order $\pi/a$.  In the tadpole-improved case,
where one hopes the most severe cutoff effects are taken out of the
perturbative renormalization parts, one might expect the values of
$q^\star$ to be much lower~\cite{Lepage93}.
Table~\ref{tb:qstarRES}  shows that, while $q^\star$ is indeed reduced
for tadpole-improved quantities, this reduction is rather minimal,
and tadpole-improved $q^\star$'s are still substantially larger than
$1/a$.

\section{Scale setting using a nonperturbative expression for the coupling}

The prescription for computing $q^\star$ given in eq.~(\ref{eq:prescription})
can be motivated starting from the assumption that a natural value for
$q^\star$ would be obtained from~\cite{Lepage93}
\begin{equation}
 \alpha( q^{\star} ) \int d^4 q f(q)   = \int d^4 q f(q)  \alpha( q )\;.
\label{eq:bestqstartDEFN}
\end{equation}
The idea here is to use
a coupling that depends on the internal momentum in the
diagram to weight the integrand, rather than multiplying the diagram
with a coupling at a constant scale.
Unfortunately
the right hand side of eq.~(\ref{eq:bestqstartDEFN}) is singular if 
the perturbative renormalization group improved evolution equation is used for
the coupling, because of the Landau singularity. Lepage and Mackenzie
avoided this problem, by using
the one loop perturbative evolution equation for the coupling
(which is free of the Landau singularity) in 
eq.~(\ref{eq:bestqstartDEFN}) to derive eq.~(\ref{eq:prescription}).

As a consistency check on the scale obtained from
eq.~(\ref{eq:prescription}), it is instructive to try to use a
nonperturbative definition of the coupling in
eq.~(\ref{eq:bestqstartDEFN}). The Landau singularity in the
renormalization group evolution equation for the coupling occurs when
the coupling is large, and the perturbative derivation of the
evolution equation breaks down.  Neubert~\cite{Neubert95}
developed a formalism based on eq.~(\ref{eq:bestqstartDEFN}) to
calculate the scale for a number of continuum perturbative
expressions. He found that his more general formalism produced well
behaved perturbative expression for some quantities that were
previously thought to be unreliable because of their low BLM scale
(the continuum analog of using eq.~(\ref{eq:prescription})).  This
motivates studying the use of eq.~(\ref{eq:bestqstartDEFN}) for
lattice perturbation theory.

As the running coupling of QCD has been obtained nonperturbatively,
from a number of lattice QCD
simulations~\cite{Michael:1992nj,Luscher:1994gh,Parrinello:1997wm}, it
seems natural to try  a coupling measured from a  lattice QCD
simulation in eq.~(\ref{eq:bestqstartDEFN}) to estimate the scale
for lattice perturbative expressions.
A useful expression for a nonperturbative QCD coupling for scale
setting was suggested by Klassen~\cite{Klassen95a}.  This QCD coupling
is defined by fitting the static potential between two quarks in
momentum space to an ansatz for the evolution of the coupling that has
no Landau pole and reduces to the perturbative expressions in the weak
coupling limit.  Klassen used eq.~(\ref{eq:bestqstartDEFN}) with
his nonperturbative coupling to estimate the scale for the plaquette,
which was consistent with the result from Lepage and Mackenzie's
formalism.

Unfortunately, combining Klassen's nonperturbative definition of the
coupling with the integrands for $Z_A$ in eq.~(\ref{eq:bestqstartDEFN})
produces a divergent integral.
The problem is that the integrands for $Z_A$ behave like 
$O(\frac{1}{q^2})$ for small momentum $q$. Klassen's coupling also 
behaves like $O(\frac{1}{q^2})$, thus producing a divergent result.
The physical reason for the low momentum limit of Klassen's coupling
is the assumption that the static quark potential at large distances is linear
in the quark separation. However, it is expected that at a  certain
separation between the quarks, string breaking will occur, which will
cause the potential to flatten out. The crossover from a linearly rising
potential to a flat potential has never been seen in lattice gauge
theory simulations (see Refs.~\cite{Bali:1997bj,Drummond:1998} for  recent 
discussions). 
>From lattice gauge theory,
it 
is thus currently impossible to extract a coupling in momentum space,
for use in eq.~(\ref{eq:bestqstartDEFN}), that 
has the correct low momentum limit.

There is some evidence from experiment that the low energy limit
of the coupling tends to a finite limit in the small momentum limit.
In the theoretical analysis of $e^+e^-$ event shapes,
Dokshitzer and Webber~\cite{Dokshitser:1995zt} introduced the low momentum mean of the
coupling.
Their fit to experimental data yielded
\begin{equation}
\frac{1}{2 \; GeV} \int_{0}^{2 \; GeV} \alpha(\mu^2) d \mu = 0.52 \pm 0.04\;.
\label{eq:loweventshape}
\end{equation}
If a low momentum coupling with the limit of eq.~(\ref{eq:loweventshape})
exists, then it would have a finite limit with the integrands for 
$Z_A$.

Inspired by the work that obtained eq.~(\ref{eq:loweventshape}),
Shirkov and Solovtsov~\cite{Shirkov:1997wi} derived an expression for the coupling
without a Landau pole:
\begin{equation}
\alpha_{ss}( k^2) = \frac{1}{ \beta_0 } 
\left(  
      \frac{ 1 } { \log( k^2/\Lambda^2  )} 
            + 
      \frac{ \Lambda^2 } { \Lambda^2 - k^2 }
\right)\;.
\label{eq:dispersiveONEloop}
\end{equation}
Equation~(\ref{eq:dispersiveONEloop}) is based on a dispersive
theory and  the one-loop RG evolution equation. The choice of
scheme is reflected in the $\Lambda$ parameter.
This definition of the coupling,
as well as its connection to renormalons,
is reviewed in Refs.~\cite{Zakharov:1997xs,Akhoury:1997hi}.
The analytic coupling in eq.~(\ref{eq:dispersiveONEloop})  has been used to
study the scale dependence in the continuum~\cite{Solovtsov:1997at}.

Although eq.~(\ref{eq:dispersiveONEloop}) could just be
substituted into the right-hand side of
eq.~(\ref{eq:bestqstartDEFN}) to provide an answer for the 
one-loop perturbative contribution, it is convenient to 
calculate a $q^\star$ to compare with the scale obtained from the 
Lepage and Mackenzie scale. We define a $q^\star$ from:
\begin{equation}
\frac{1}{\log \left( (q^\star)^2/  \lambda^2 \right)}  + 
\frac{1}{1-(q^\star)^2/\lambda^2}
= 
\frac
    { \int d^4 q f(q) \left( 1/\log( \frac{q^2}{ \lambda^2} )  +
\frac{1}{ 1 - q^2/ \lambda^2  } \right)} 
    { \int d^4 q f(q) }\;, 
\label{eq:coupling} 
\end{equation}
where $q$ and $\lambda \equiv a\Lambda$ are dimensionless.
Using $\lambda = 0.14$ ($\lambda = 0.0587$), taken from 
Ref.~\cite{Klassen95a}  at
$\beta = 6.0$ ($\beta = 6.8$),
we obtain $q^\star = 2.32$  ($q^\star = 2.33$) for  $Z_A$, 
and $q^\star = 2.10$ ($q^\star = 2.10$) for the 
tadpole improved (using  $M_c$) $Z_A$.
The two scales are both close to 
ones obtained from the Lepage and
Mackenzie scheme. Again we see that tadpole improving the 
perturbative factor only causes a small reduction in the scale.
Although eq.~(\ref{eq:coupling}) is less convenient to use 
with simulation data, because of the dependence on the 
scale through $\lambda$, this formalism does provide a check on the 
Lepage and Mackenzie scale-setting procedure.

\section{Visualization of the integrands}

To try to understand why the estimate of the scale 
for tadpole improved perturbative factors was not close to the naive
expectation $\frac{1}{a}$, and to investigate the
effects of having different weighting functions, we wanted to 
``visualize'' the lattice integrands. Consider a lattice 
approximation to a continuum integral
\begin{eqnarray}
\int^{2 \pi }_{0} f(q) d^4q & \sim  &
   \frac{ 1} { L^4} \sum_{x=0}^{L-1} \sum_{y=0}^{L-1} \sum_{z=0}^{L-1} \sum_{t=0}^{L-1} 
   f( 2 \pi x/L , 2\pi y /L  , 2 \pi z/L,2 \pi t/L  ) \\
& = & \frac{ 1} { L^4}  \sum_{ q_E } f(q_E) n(q_E)\;,
\end{eqnarray}
where $q_E$ is a single member of the equivalence class under hypercubic
transformations. The function  $n(q_E)$ gives the number of momenta
in each equivalence class. This technique is the basis of an efficient
method to integrate lattice Feynman diagrams~\cite{Luscher86a}.

\begin{figure}[t]
\def\filename{za_tadpole_yes_and_no_small.ps}
\begin{center}
\epsfysize=\hsize \leavevmode 
\epsfbox{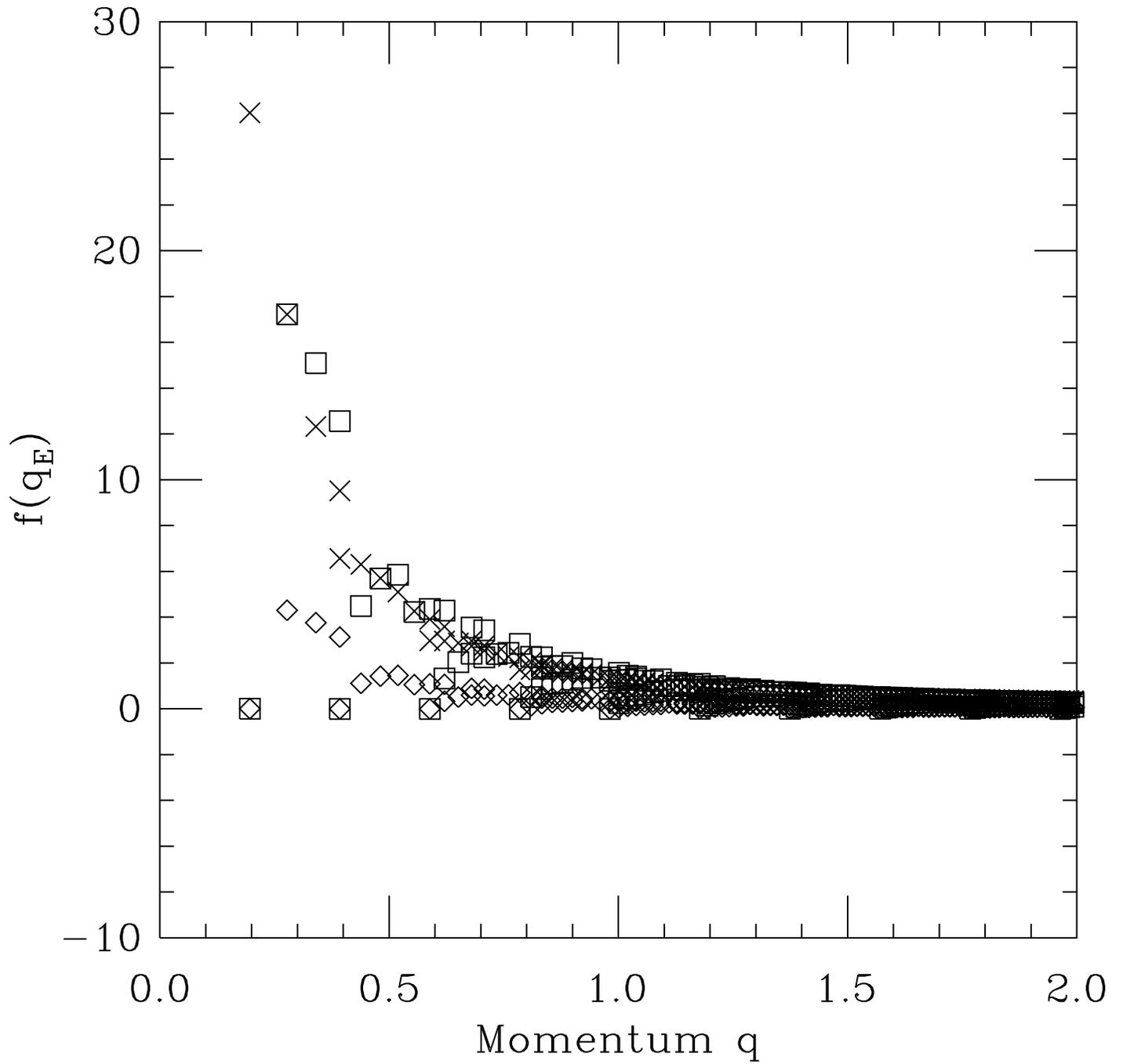}
\end{center}
  \caption{Plots of the integrand for $Z_A$ with (diamonds)  and without (crosses)
tadpole 
improvement (through \protect{$M_c$} ). The squares are the tadpole improved 
integrand normalized to
agree with the nontadpole improved numbers at a specific point. The
lattice volume was \protect{$32^4$}. For clarity only momentum with
magnitude less than two are displayed.
 }
\label{fig:zaplot}
\end{figure}

In figure~\ref{fig:zaplot}, we plot the $f(q_E)$ function for the $Z_A$
integrand with and without tadpole improvement. The magnitude of the 
tadpole-improved integrand is much reduced over the original
integrand. However, if we choose the normalization of the 
tadpole-improved integrand to equal the not improved
integrand at a specific momentum, then the two figures agree 
qualitatively. Because the Lepage-Mackenzie scale does not depend
on the overall normalization of the integrands, it is perhaps 
not surprising that the scales for the tadpole and nontadpole
$Z_A$ are so close.

\section{Conclusion} \label{Sec:Con}

Our main results for the perturbative scales are contained in
table~\ref{tb:qstarRES}. Tadpole improvement does not significantly
reduce $q^\star$ for the operators considered here. In particular,
the scales for the tadpole-improved operators are higher than the 
scale $q^\star = 1/a$  suggested by the intuitive idea that tadpole
improvement reduces the effect of the momentum near the cut 
off~\cite{Lepage93}.

The closeness of the scale for the tadpole-improved  $Z_A$ to the 
scale for the tadpole-improved $Z_A$ in the static limit~\cite{Hernandez94a} 
($aq^\star = 2.18$)
seems tp indicate that the mass dependence of $q^\star$ is weak. This suggests
that using a mass-independent $q^\star$ with the results of the mass-dependent 
perturbative calculation of $Z_A$ by Kuramashi~\cite{Kuramashi1997} is a good 
approximation.
This was the procedure adopted in the calculation of $f_B$
by the MILC collaboration~\cite{BernardfB98}.

\acknowledgements

We thank Urs Heller for discussions about
the low-energy limit of the coupling and string breaking.
This work is supported in part by the U.S.\ Department of Energy
under grant number DE-FG02-91ER40628
and the NSF under grant numbers NSF PHY 96-01227 and NSF ASC 89-0282.

\pagebreak[4]

%%
%%  Tables section
%%

%
\begin{table}[t]
\begin{center}
\begin{tabular}{c||c|c|c} \hline
& {\rm no improvement} &
{\rm Landau gauge }& {\rm through $M_c$} \\ \hline
      &          &          &          \\
$c_A$ & 0.133373(2) & 0.055907(1) & 0.024803(1) \\ \hline
      &          &          &          \\
$c_V$ & 0.174083(3) & 0.096617(2) & 0.065512(1)\\ \hline
      &          &          &          \\
$c_{mP}$ & 0.081419(2) & -- & -- \\ \hline
\end{tabular}
\end{center}
\caption{ One-loop constants for tadpole-improved
renormalization constants. }
\vspace{0.2cm}
\label{tb:perturbcontants}
\end{table}
\begin{table}[t]
\begin{center}
\begin{tabular}{c||c|c|c} \hline
{\rm quantity} &
{\rm no improvement }& {\rm Landau gauge } & {\rm through $M_c$ } \\ \hline
 &  &   &   \\
$Z_A$ & 2.533 & 2.235 & 2.316 \\
% & &  &  \\
\hline
 & &  &   \\
$Z_V$ & 2.370 & 2.090 & 2.051 \\
% & &   &   \\
\hline
 &  &  &   \\
$Z_{mP}$ & 1.905 &  &  \\
% &  &  &  \\
\hline
\end{tabular}
\end{center}
\caption{ The scale $q^\star$ (in units of $1/a$) for various quantities and improvement
schemes. The errors are less than 1 in the last place.}
\vspace{0.2cm}
\label{tb:qstarRES}
\end{table}
%

%%%
%%%  figures section
%%%

\end{document}